\documentclass[twocolumn,twoside,slac_two]{revtex4}
\usepackage{graphicx}
\usepackage{fancyhdr}
\begin{document}

\title{$B^0_s$ mixing and decays at the Tevatron}

\author{Mossadek Talby (On behalf of the D0 and CDF Collaborations)}
\affiliation{CPPM, IN2P3-CNRS, Universit\'e de la M\'editerran\'ee, Marseille, France}

\begin{abstract}
This short review reports on recent results from CDF and D\O\ experiments at the Tevatron collider on 
$B^0_s$ mixing and the lifetimes of $B^0_s$ and $\Lambda_b$.
\end{abstract}

\maketitle

\thispagestyle{fancy}

\section{Introduction}
Due to the large $b\bar{b}$ cross section at 1.96 TeV $p\bar{p}$ collisions, the Tevatron collider at Fermilalb 
is currently the largest source of $b$-hadrons and provides a very rich environment for the study of $b$-hadrons. 
It is also the unique place to study high mass $b$-hadrons such as $B^0_s$, $B_c$, $b$-baryons and excited $b$-hadrons states. \par

CDF and D\O\ are both symmetric multipurpose detectors~\cite{CDFdet,D0det}. They are essentially similar and consist of vertex 
detectors, high resolution tracking chambers in a magnetic field, finely segmented hermitic calorimeters and muons momentum 
spectrometers, both providing a good lepton identification. They have fast data acquisition systems with several levels of online 
triggers and filters and are able to trigger at the hardware level on large track impact parameters, enhancing the potential of 
their physics programs. 

\section{$B^0_s$ mixing}
The $B^0$-$\bar{B}^0$ mixing is a well established phenomenon in particle physics. It proceeds via a flavor changing weak 
interaction in which the flavor eigenstates $B^0$ and $\bar{B}^0$ are quantum superpositions of the two mass eigenstates $B_H$ 
and $B_L$. The probability for a $B^0$ meson produced at time $t=0$ to decay as $B^0$ or $\bar{B}^0$ at proper 
time $t>0$ is an oscillatory function with a frequency $\Delta m$, the difference in mass between $B_H$ and $B_L$. 
Oscillation in the $B^0_d$ system is well established experimentally with a precisely measured oscillation frequency 
$\Delta m_d$. The world average value is $\Delta m_d = 0.507\pm 0.005$ ps$^{-1}$~\cite{dmd}. In the $B^0_s$ system, the expected 
oscillation frequency value within the standard model (SM) is approximately 35 times faster than $\Delta m_d$. In the SM, 
the oscillation frequencies $\Delta m_d$ and $\Delta m_s$ are proportional to the fundamental CKM matrix elements $|V_{td}|$ 
and $|V_{ts}|$ respectively, and can be used to determine their values. This determination, however, has large theoretical 
uncertainties, but the combination of the $\Delta m_s$ measurement with the precisely measured $\Delta m_d$ allows the determination  
of the ratio $|V_{td}|/|V_{ts}|$ with a significantly smaller theoretical uncertainty.  \par 

Both D\O\ and CDF have performed $B^0_s$-$\bar{B}^0_s$ mixing analysis using 1 fb$^{-1}$ of data~\cite{D0Bmix,CDFBmix1,CDFBmix2}. 
The strategies used by the two experiments to measure $\Delta m_s$ are very similar. They schematically proceed as follows: 
the $B^0_s$ decay is reconstructed in one side of the event and its flavor at decay time is determined from its decay products. 
The $B^0_s$ proper decay time is measured from the the difference between the $B^0_s$ vertex and the primary vertex of the event. 
The $B^0_s$ flavor at production time is determined from information in the opposite and/or the same-side of the event. finally, 
$\Delta m_s$ is extracted from an unbinned maximum likelihood fit of mixed and unmixed events, which combines, among other 
information, the decay time, the decay time resolution and $b$-hadron flavor tagging. In the following only the latest CDF 
result is presented. \par

\subsection{$B^0_s$ signal yields}
\begin{figure}[h]
\centering
\includegraphics[width=80mm,height=75mm]{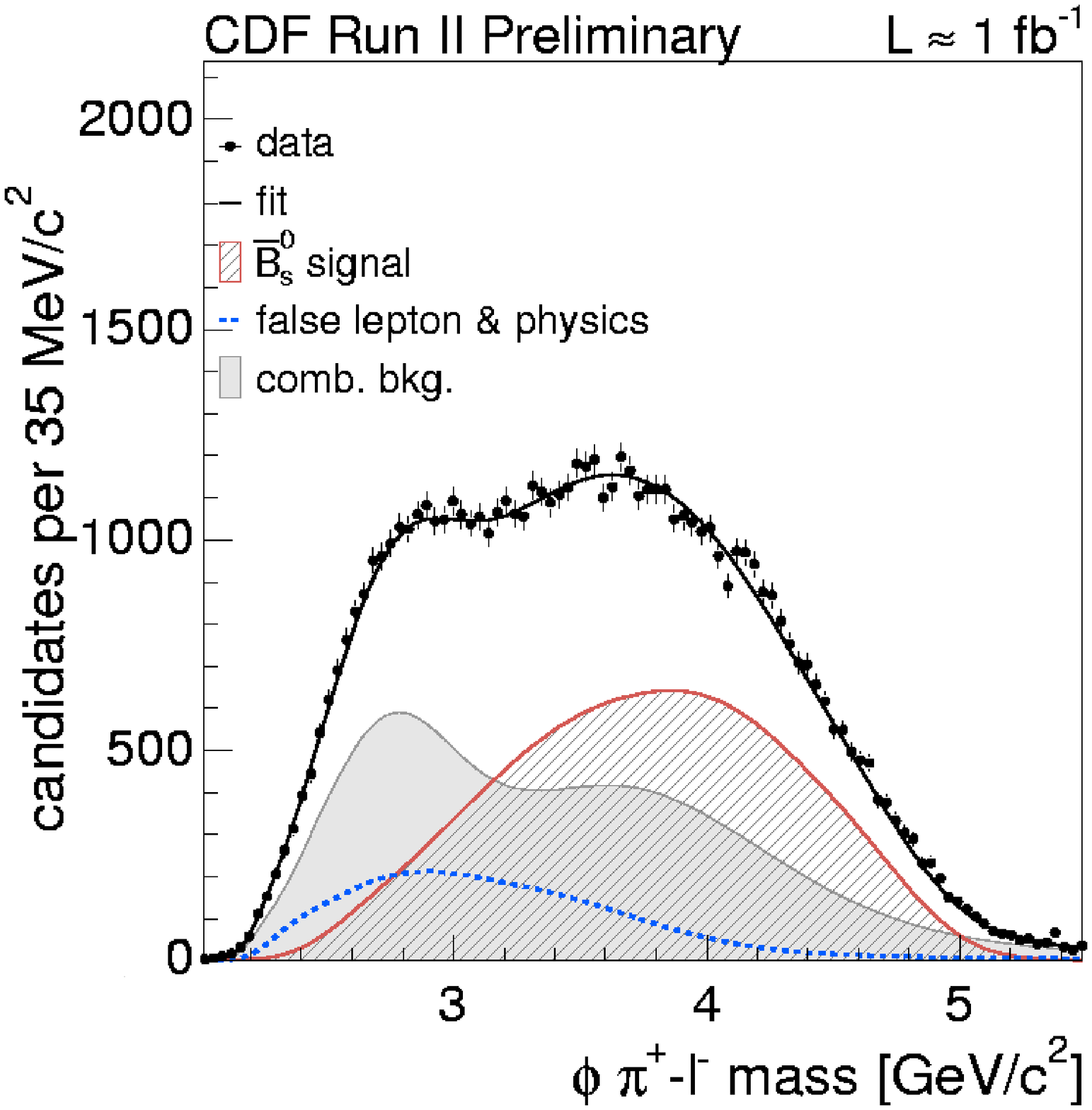}
\includegraphics[width=80mm,height=75mm]{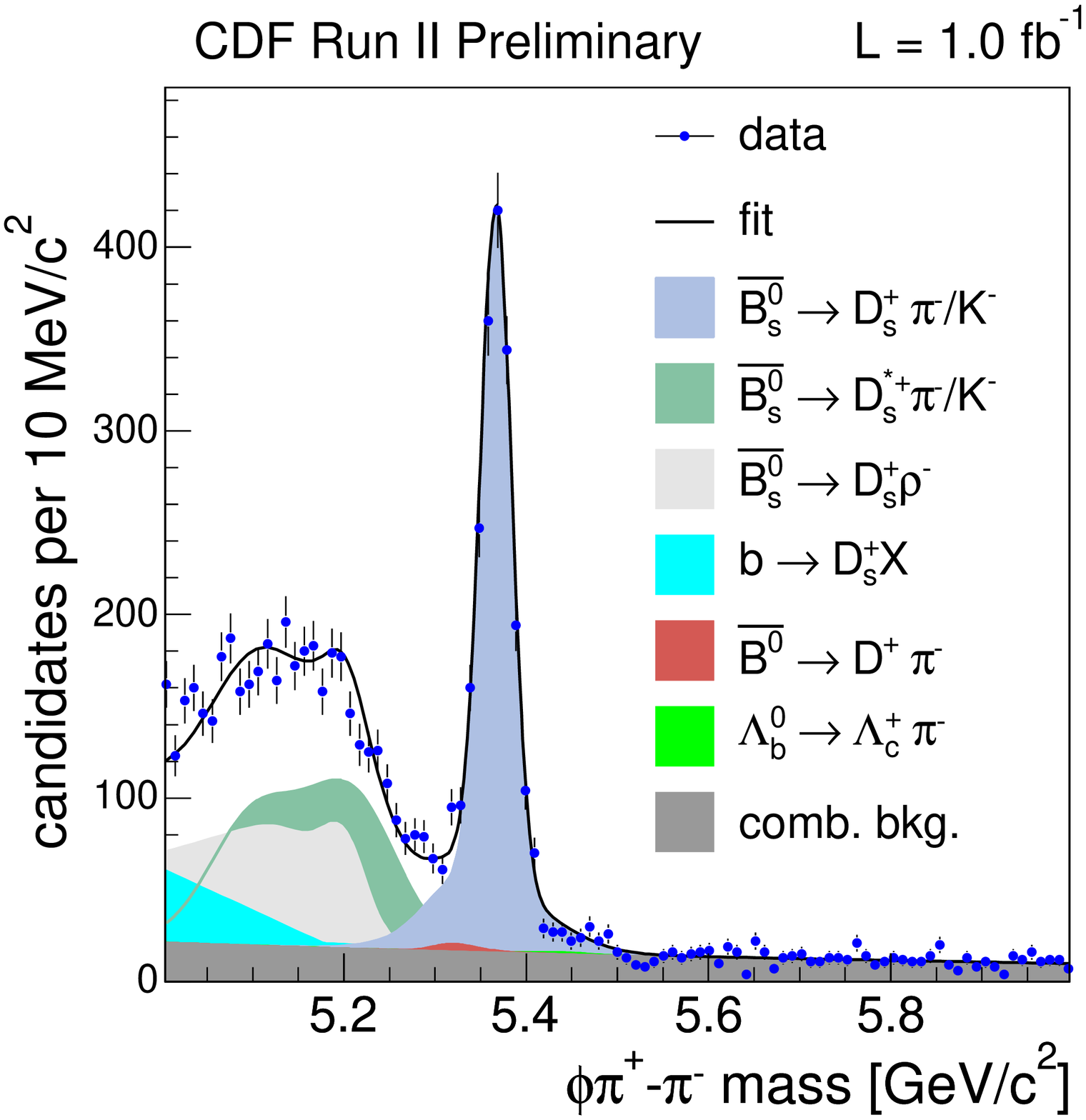}
\caption{The invariant mass distributions for the $D^+_s(\phi\pi^+)\ell^-$ pairs (upper plot) and for the 
$\bar{B}^0_s\rightarrow D^+_s(\phi\pi^+)\pi^-$ decays (bottom plot) including the contributions from the partially reconstructed hadronic decays.} 
\label{Bsreco}
\end{figure}

The CDF experiment has reconstructed $B^0_s$ events in both semileptonic $B^0_s\rightarrow D^{-(\star)}_s \ell^+\nu_{\ell} X$ 
($\ell = e$ or $\mu$) and hadronic $B^0_s\rightarrow D^-_s(\pi^+\pi^-)\pi^+$ decays. In both cases the $D^-_s$ is reconstructed 
in the channels $D^-_s\rightarrow \phi\pi^-$, $D^-_s\rightarrow K^{\star 0} K^-$ and $D^-_s\rightarrow\pi^-\pi^+\pi^-$ with 
$\phi\rightarrow K^+K^-$ and $K^{\star 0}\rightarrow K^+\pi^-$. Additional partially reconstructed hadronic decays, 
$B^0_s\rightarrow D^{\star -}_s\pi^+$ and $B^0_s\rightarrow D^{-}_s\rho^+$ with unreconstructed $\gamma$ and $\pi^0$ in 
$D^{\star -}_s\rightarrow D^-_s(\phi\pi^-)\gamma /\pi^0$ and  $\rho^+\rightarrow \pi^+\pi^0$ decay modes, have also been used. 
The signal yields are 61,500 semileptonic decays, 5,600 fully reconstructed and 3,100 partially reconstructed hadronic decays. 
This correponds to an effective statistical increase in the number of reconstructed events of 2.5 compared to the first CDF 
published analysis~\cite{CDFBmix1}. This improvement was obtained mainly  by using particle identification in the event selection, 
by using the artificial neural network (ANN) selection for hadronic modes and by loosening the kinematical selection. 
Figure~\ref{Bsreco} shows the distributions of the invariant masses of the $D^+_s(\phi\pi^+)\ell^-$ pairs $m_{D_s\ell}$ 
and of the $\bar{B}^0_s\rightarrow D^+_s(\phi\pi^+)\pi^-$ decays including the contributions from the partially reconstructed 
hadronic decays. \par

\subsection{$B^0_s$ proper decay time reconstruction}
\begin{figure}[h]
\centering
\includegraphics[width=80mm,height=75mm]{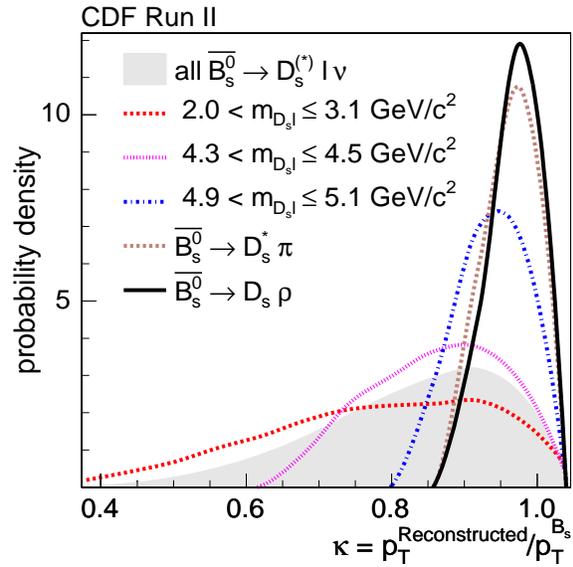}
\caption{The distribution of the correction factor $K$ in semileptonic and partially reconstructed hadronic decays from Monte Carlo simulation.} \label{Kfactor}
\end{figure}

The proper decay time of the reconstructed $B^0_s$ events is determined from the transverse decay length $L_{xy}$ 
which corresponds to the distance between the primary vertex and the reconstructed $B^0_s$ vertex projected onto the transverse 
plane to the beam axis. For the fully reconstructed $B^0_s$ decay channels the proper decay time is well defined and is given by: 
\begin{eqnarray*} 
t = L_{xy}\frac{M(B^0_s)}{P_T(B^0_s)}
\end{eqnarray*} 
For the partially reconstructed $B^0_s$ decay channels it is given by:
\begin{eqnarray*} 
t = L_{xy}\frac{M(B^0_s)}{P_T(D_s\ell(\pi))}\times K,\;\;\; K=\frac{P_T(D_s\ell(\pi))}{P_T(B^0_s)}
\end{eqnarray*} 
The $K$-factor takes into account the missing particles\footnote{Neutrino, $\pi^0$ and $\gamma$.} in the event. It's distribution 
for different $B^0_s$ decay channels is obtained from Monte Carlo simulation. For illustration, figure~\ref{Kfactor} shows 
the $K$-factor distributions obtained by CDF in semileptonic and partially reconstructed hadronic decays. 
The proper time resolution which is one of the critical parameters for $\Delta m_s$ measurement, has contributions from the 
uncertainty on $L_{xy}$ and from the momentum of the missing decay products. For the fully reconstructed decay modes, 
the mean proper decay time resolution obtained is 87 fs, which corresponds to the quarter of an oscillation period at 
$\Delta m_s=17.8$ ps$^{-1}$. For the partially reconstructed hadronic and semileptonic decays, the average proper decay 
time resolutions are $\sigma_t=97$ fs and $\sigma_t=168$ fs respectively. \par

\subsection{Flavor tagging}
The flavor of the $B^0_s$ at the decay time is determined precisely from its decay products. At production time, the flavor of 
the $B^0_s$ is determined using both opposite-side and the same-side $b$-flavor tagging techniques. 
The opposite-side tagging exploits the fact that $b$ quarks are dominently produced in $b\bar{b}$ pairs in hadron colliders. 
Same side tagging relies on the charges and the identity of associated particles produced in the fragmentation of the $b$ quark 
that produces the reconstructed $B^0_s$. The effectiveness, $Q\equiv\epsilon\mathcal D^{2}$, of these techniques is quantified 
with an efficiency $\epsilon$, the fraction of signal candidates with a flavor tag, and a dilution $\mathcal D=1-2\omega$, 
where $\omega$ is the probablity of mistagging. \par 

The taggers used in the opposite-side of the event are the charge of the lepton ($e$ and $\mu$), the jet charge and the charge of 
identified kaons. The information from these taggers are combined in an ANN. The use of an ANN improves the combined 
opposite-side tag effectiveness by 20\% ($Q=1.8\pm 0.1$\%) compared to the previous analysis~\cite{CDFBmix1}. The dilution is measured 
in data using large samples of kinematically similar $B^0_d$ and $B^+$ decays. \par

The same-side flavor tags rely on the identification of the charge of the kaon produced from the left over $\bar{s}$ in the 
process of $B^0_s$ fragmentation. Any nearby charged particle to the reconstructed $B^0_s$, identified as a kaon, is expected 
to be correlated to the $B^0_s$ flavor, with a $K^+$ correlated to a $B^0_s$ and $K^-$ correlated to $\bar{B}_s$. An ANN is used 
to combine particle-identification likelihood based on information from the $dE/dx$ and from the Time-of-Flight system, with 
kinematic quantities of the kaon candidate into a single variable. The dilution of the same side tag is estimated using Monte 
Carlo simulated data samples. The predicted effectiveness of the same-side flavor tag is Q=3.7\% (4.8\%) in the hadronic 
(semileptonic) decay sample. The use of ANN increased the $Q$ value by 10\% compared to the previous analysis~\cite{CDFBmix1}. \par
If both a same-side tag and an opposite-side tag are present, the information from both tags are combined assuming they are 
independent. \par

\subsection{$\Delta m_s$ measurement}
\begin{figure}[h]
\centering
\includegraphics[width=80mm,height=65mm]{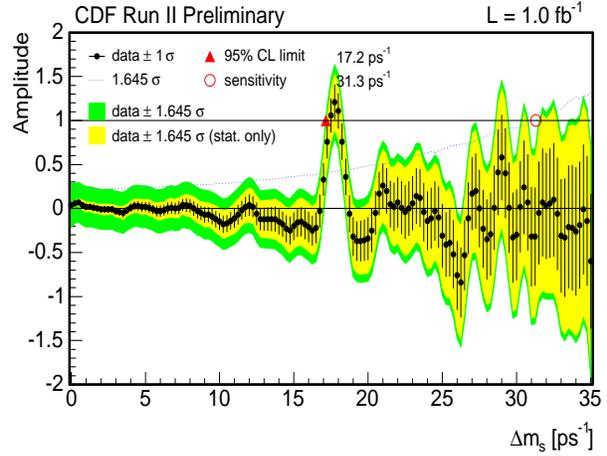}
\caption{The measured amplitude values and uncertainties versus the $B^0_s$-$\bar{B}^0_s$ oscillation frequency $\Delta m_s$.} \label{ampscan}
\end{figure}

An unbinned maximum likelihood fit is used to search for $B^0_s$ oscillations in the reconstructed $B^0_s$ decays samples. 
The likelihood combines masses, decay time, decay-time rsolution, and flavor tagging information for each reconstructed $B^0_s$ 
candidate, and includes terms for signal and each type of background. The technique used to extract $\Delta m_s$ 
from the unbinned maximum likelihood fit, is the amplitude scan method~\cite{Moser} which consists of multiplying the oscillation 
term of the signal probablity density function in the likelihood by an amplitude $\mathcal A$, and fit its value for different 
$\Delta m_s$ values. The oscillation amplitude is expected to be consistent with $\mathcal A = 1$ when the probe value is the 
true oscillation frequency, and consitent with $\mathcal A = 0$ when the probe value is far from the true oscillation frequency. 
Figure~\ref{ampscan} shows the fitted value of the amplitude as function of the oscillation frequency for the combination of 
semileptonic and hadronic $B^0_s$ candidates. \par
\begin{figure}[h]
\centering
\includegraphics[width=80mm]{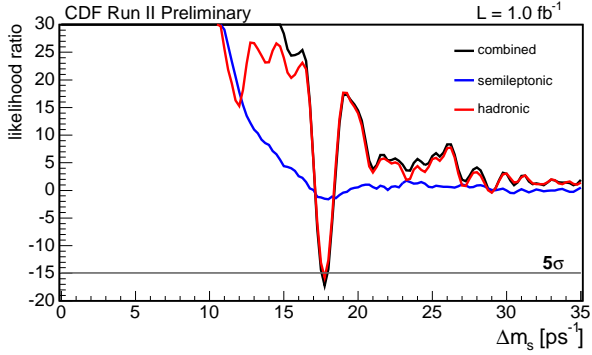}
\caption{The logarithm of the ratio of likelihoods 
$\Lambda \equiv \mathrm{log}\left[\mathcal L^{\mathcal A=0}/\mathcal L^{\mathcal A=1}(\Delta m_s)\right]$, versus 
the oscillation frequency. The horizontal line indicates the value $\Lambda=-15$ that corresponds to a probability of 
$5.7\times 10^{-7}$ (5$\sigma$) in the case of randomly tagged data.} \label{loglikelihood}
\end{figure}

The sensitivity is 31.3 ps$^{-1}$ for the combination\footnote{19.3 ps$^{-1}$ for the semileptonic decays alone and 30.7 ps$^{-1}$ 
for the hadronic decays alone.} of all hadronic and semileptonic decay modes. At $\Delta m_s=17.75$ ps$^{-1}$, the observed 
amplitude $\mathcal A = 1.21\pm 0.20$ (stat.) is consistent with unity and $\mathcal A/\sigma_{\mathcal A}= 6.05$ where 
$\sigma_{\mathcal A}$ is the statistical uncertainty on $\mathcal A$. This shows that the amplitude is inconsistent with zero 
and that the data are compatible with $B^0_s$-$\bar{B}^0_s$ oscillations with that frequency. The significance of the signal is 
evaluated using the logarithm likelihood ratio 
$\Lambda \equiv \mathrm{log}\left[\mathcal L^{\mathcal A=0}/\mathcal L^{\mathcal A=1}(\Delta m_s)\right]$. 
Figure~\ref{loglikelihood} shows $\Lambda$ as function of $\Delta m_s$. At the minimum $\Delta m_s=17.77$ ps$^{-1}$, 
$\Lambda=-17.26$. The significance of the signal is the probability that randomly tagged data would produce a value of 
$\Lambda$ lower than -17.26 at any $\Delta m_s$ value. This probability has been determined to be $8\times 10^{-8}$ which 
correponds to a significance of 5.4 $\sigma$. \par

The $\Delta m_s$ value is determined from the fit for oscillation frequency at amplitude value  $\mathcal A=1$. The fit result 
is $\Delta m_s =17.77\pm 0.10\; (\mathrm{stat.})\pm 0.07\; (\mathrm{sys.})$ ps$^{-1}$. The systematic error is completely 
dominated by the time scale uncertainty. \par
The measured $B^0_s$-$\bar{B}^0_s$ oscillation frequency is used to determine the ratio $|V_{td}/V_{ts}|$. If one uses as inputs 
$m_{B^0_d}/m_{B^0_s}=0.98390$~\cite{Bmass_ratio} with negligeable uncertainty, $\Delta m_d=0.507\pm 0.005$ ps$^{-1}$~\cite{dmd} and 
$\xi=1.21^{+0.047}_{-0.035}$~\cite{csi}, one finds:
\begin{eqnarray*} 
|V_{td}/V_{ts}|&=&\xi\sqrt{\frac{\Delta m_d m_{B^0_s}}{\Delta m_s m_{B^0_d}}}\\
&=& 0.2060\pm0.0007(\mathrm{exp})^{+0.0081}_{-0.006}(\mathrm{theor})
\end{eqnarray*} 
\section{$b$-hadrons lifetime measurements at the Tevatron RunII}
Lifetime measurements of $b$-hadrons provide important information on the interactions between heavy and light quarks. These 
interactions are responsible for lifetime hierarchy among $b$-hadrons observed experimentally: 
\begin{eqnarray*} 
\tau(B^+)\geq\tau(B^0_d)\simeq\tau(B^0_s)>\tau(\Lambda_b)\gg\tau(B_c)
\end{eqnarray*} 
Currently most of the theoretical calculations of the light quark effects on $b$ hadrons lifetimes are performed in the framework 
of the Heavy Quark Expansion (HQE)~\cite{HQE} in which the decay rate of heavy hadron to an inclusive final state $f$ is expressed 
as an expansion in $\Lambda_{\mathrm{QCD}}/m_b$. At leading order of the expansion, light quarks are considered as spectators and 
all $b$ hadrons have the same lifetime. Differences between meson and baryon lifetimes arise at 
${\mathcal O} (\Lambda^2_{\mathrm{QCD}}/m^2_b)$ and splitting of the meson lifetimes appears at 
${\mathcal O} (\Lambda^3_{\mathrm{QCD}}/m^3_b)$. \par

Both CDF and D\O\ have performed a number of $b$-hadrons lifetimes measurements for all $b$-hadrons species. Most of these 
measurements are already included in the world averages and are summarised in~\cite{HFAG07}. In this note focus will be on 
the latest results on $B^0_s$ and $\Lambda_b$ measurements from CDF and D\O\ . \par
 
\subsection{$B^0_s$ lifetime measurements} \par
In the standard model the light $B_{L}$ and the heavy $B_{H}$ mass eigenstates of the mixed $B^0_s$ system are expected 
to have a sizebale decay width difference of order $\Delta\Gamma_s= \Gamma_L-\Gamma_H = 0.096\pm 0.039$ 
ps$^{-1}$~\cite{Dgamma}. If CP violation is neglected, the two $B^0_s$ mass eigenstates are expected to be CP eigenstates, 
with $B_{L}$ being the CP even state and $B_{H}$ being the CP odd state. \par

Various $B^0_s$ decay channels have a different proportion of $B_{L}$ and $B_{H}$ eigenstates:
\begin{itemize}
\item Flavor specific decays, such as $B^0_s\rightarrow D^+_s\ell^-\bar{\nu}_{\ell}$ and 
$B^0_s\rightarrow D^+_s\pi^-$ have equal fractions of $B_{L}$ and $B_{H}$ at $t=0$. The fit to 
the proper decay length distributions of these decays with a single signal exponential lead to a flavor 
specific lifetime: 
\begin{eqnarray*} 
\tau_{B_s}(fs)=\frac{1}{\Gamma_s}\frac{1+\left(\frac{\Delta\Gamma_s}{2\Gamma_s}\right)^2}{1-\left(\frac{\Delta\Gamma_s}{2\Gamma_s}\right)^2},\;\;\; \Gamma_s=\frac{\Gamma_L+\Gamma_H}{2}
\end{eqnarray*} 
\item Fully exculusive $B^0_s\rightarrow J/\psi\phi$ decays are expected to be dominated by CP even state and its lifetime.
\end{itemize}

\subsubsection{$B^0_s$ lifetime measurements in flavor specific modes} 
Both CDF and D\O\ have measured $B^0_s$ lifetime in the semileptonic decays $B^0_s\rightarrow D^+_s\ell^-\bar{\nu}_{\ell}$. 
Results based on respectively 360 and 400 pb$^{-1}$ were presented at the FPCP06 conference last year~\cite{BsFsD0,BsFsCDF1}. 
These are:
\begin{eqnarray*} 
\tau^{\mathrm{D\O}}_{B_s}(fs) &=& 1.398\pm 0.044(\mathrm{stat})^{+0.028}_{-0.025}(\mathrm{sys})\; \mathrm{ps} \\
\tau^{\mathrm{CDF}}_{B_s}(fs) &=& 1.381\pm 0.055(\mathrm{stat})^{+0.052}_{-0.046}(\mathrm{sys})\; \mathrm{ps}.
\end{eqnarray*} 
\noindent
The D\O\ measurement provides the world best $B^0_s$ lifetimes measurement in the flavor specific decays. \par

\begin{figure}[h]
\centering
\includegraphics[width=80mm,height=70mm]{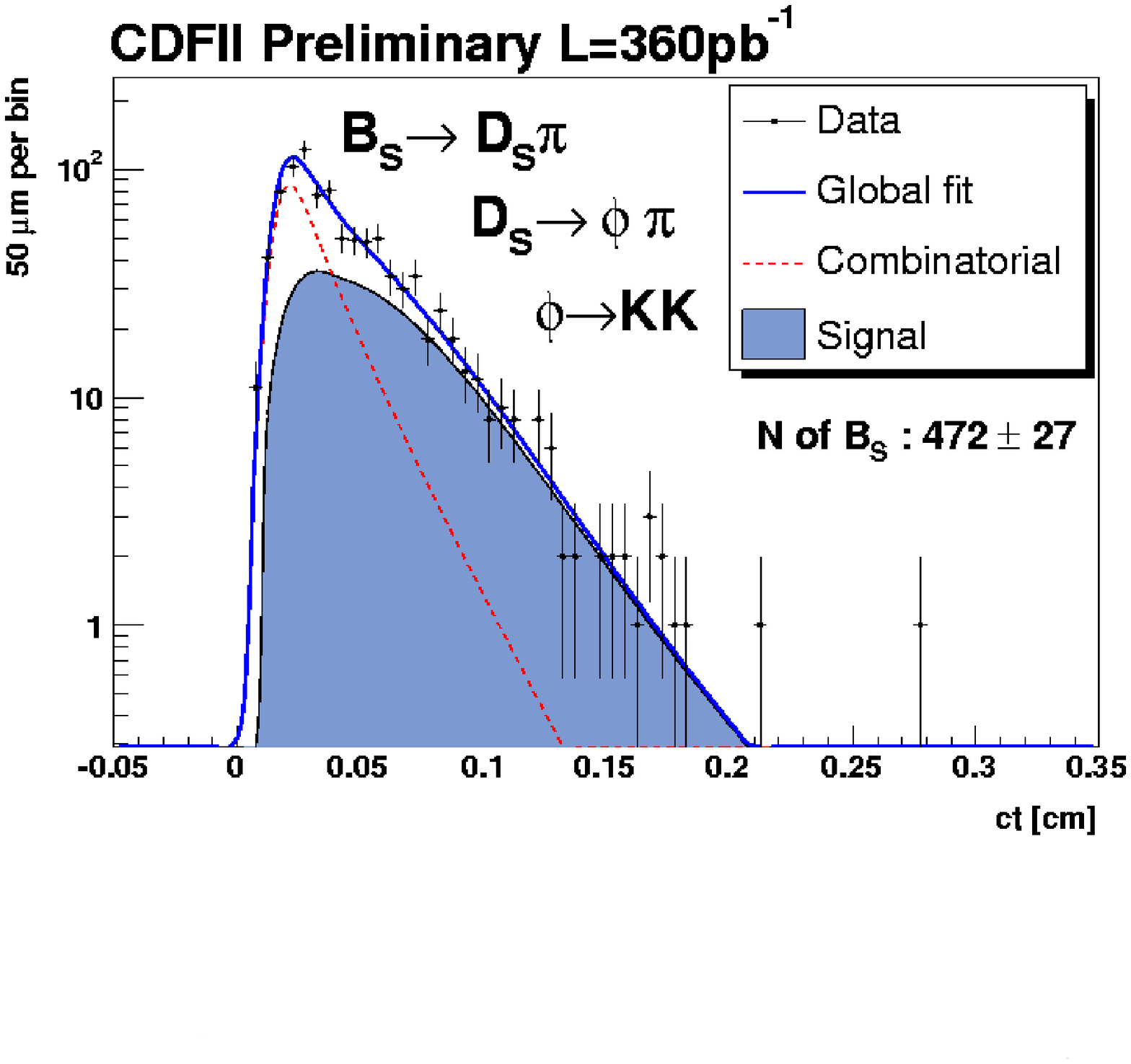}
\vskip -1.8cm
\caption{Proper decay length distribution of the $B^0_s$ candidates, with the fit result superimposed. 
The shaded region represents the signal.} \label{Bslifehad}
\end{figure}

CDF has also measured $B^0_s$ lifetime in the fully hadronic modes, $B^0_s\rightarrow D^+_s\pi^-$ and 
$B^0_s\rightarrow D^+_s\pi^+\pi^-\pi^-$. To date the analysis is based only on 360 pb$^{-1}$. The $B^0_s$ lifetime is extracted 
from a simultaneous fit to the mass and decay length distributions of the two decay modes. Figure~\ref{Bslifehad} shows the 
distribution of the proper decay length and fits to the $B^0_s$ candidates. The measured $B^0_s$ lifetime is~\cite{BsFsCDF2}
$\tau^{\mathrm{CDF}}_{B_s}= 1.60\pm 0.10(\mathrm{stat})\pm 0.02(\mathrm{sys})$ ps. These measurements in the 
semileptonic and the hadronic decay modes will soon be updated with 1 fb$^{-1}$ of data. \par 

\subsubsection{$B^0_s$ lifetime measurements in $B^0_s\rightarrow J/\psi\phi$} 
\begin{figure}[h]
\centering
\includegraphics[width=80mm,height=65mm]{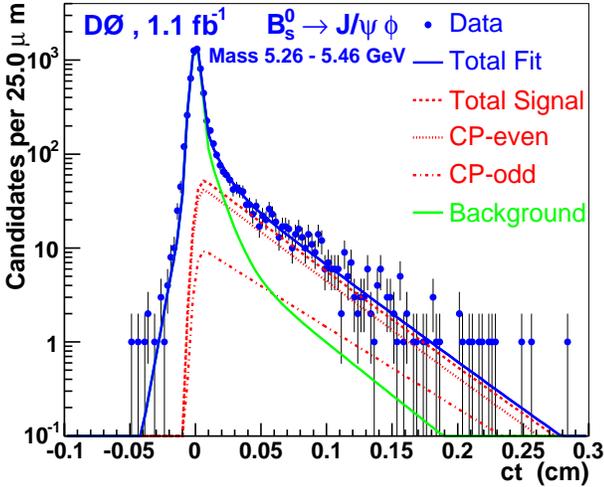}
\caption{The proper decay length, $ct$, of the $B^0_s$ candidates in the signal mass region. The curves show: the signal contribution, dashed (red); 
the CP-even (dotted) and CP-odd (dashed-dotted) contributions of the signal, the background, light solid (green); and total, solid (blue).} \label{BsJpsi}
\end{figure}

D\O\ experiment has performed a new $B^0_s$ mean lifetime measurement in $B^0_s\rightarrow J/\psi\phi$ decay mode. 
The analysis uses a data set of 1.1 fb$^{-1}$ and extracts three parameters, the average $B^0_s$ lifetime 
$\bar{\tau}(B^0_s)=1/\Gamma_s$, the width difference between the $B^0_s$ mass eigenstates $\Delta\Gamma_s$ and 
the CP-violating phase $\phi_s$, through a study of time-dependent angular distribution of the decay products 
of the $J/\psi$ and $\phi$ mesons. Figure~\ref{BsJpsi} shows the distribution of the proper decay length and fits 
to the $B^0_s$ candidates. From a fit to the CP-conserving time-dependent angular distributions of untagged decay 
$B^0_s\rightarrow J/\psi\phi$, the measured values of the average lifetime of the $B^0_s$ system and the width difference 
between the two $B^0_s$ mass eigenstates are~\cite{BsJpsiphi}:
\begin{eqnarray*} 
\tau^{\mathrm{D\O}}_{B_s} &=& 1.52\pm 0.08(\mathrm{stat})^{+0.01}_{-0.03}(\mathrm{sys}) \; \mathrm{ps} \\
\Delta\Gamma_s &=& 0.12^{+0.08}_{-0.10}(\mathrm{stat})\pm 0.02(\mathrm{sys})\;  \mathrm{ps}^{-1} \\
\end{eqnarray*} 
Allowing for CP-violation in $B^0_s$ mixing, D\O\ provides the first direct constraint on the CP-violating phase, 
$\phi_s = -0.79\pm 0.56(\mathrm{stat})^{+0.14}_{-0.01}(\mathrm{sys})$, value compatible with the standard model expectations. \par

\subsection{$\Lambda_b$ lifetime measurements} 
\begin{figure}[h]
\centering
\includegraphics[width=80mm]{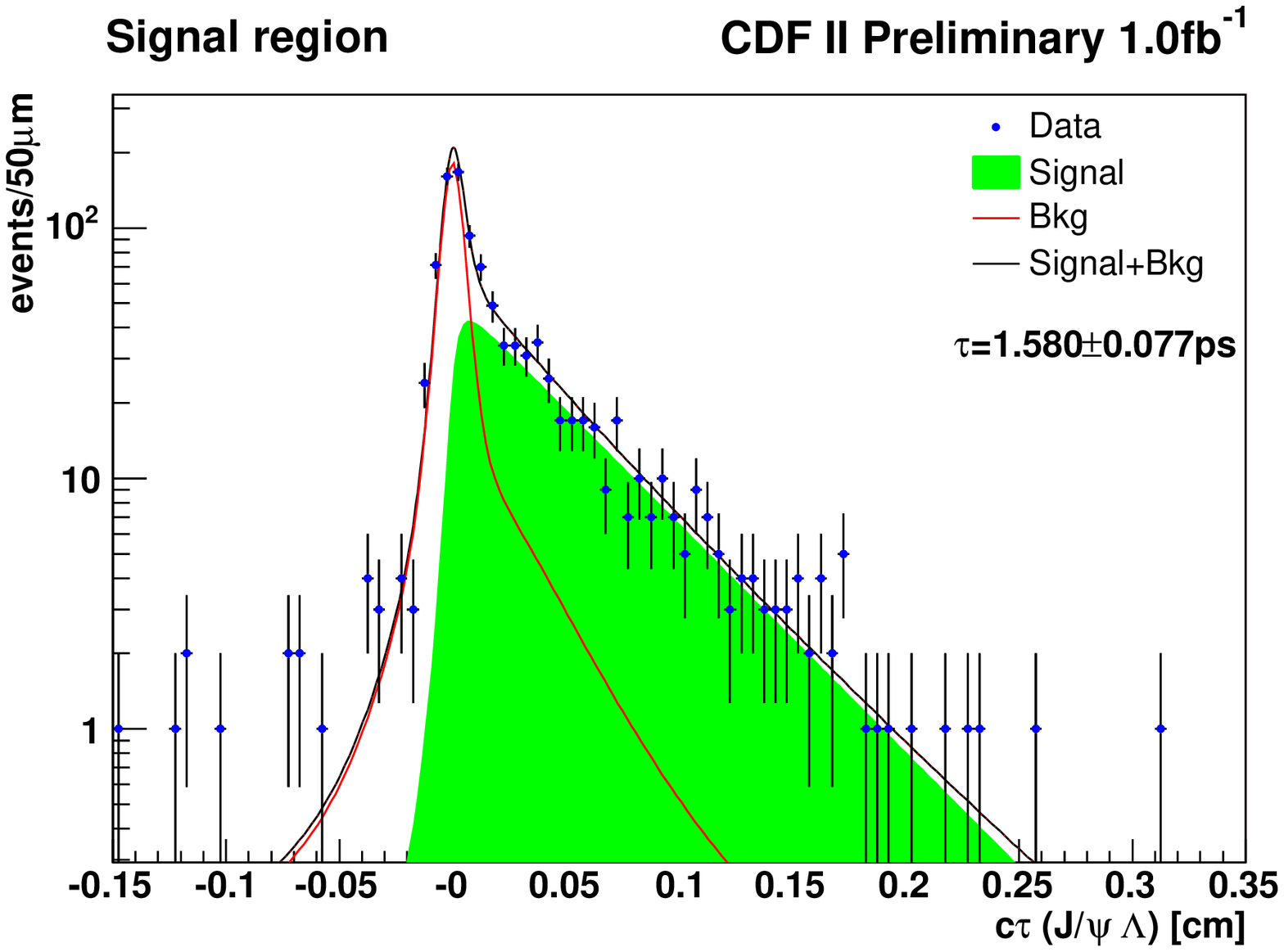}
\includegraphics[width=80mm]{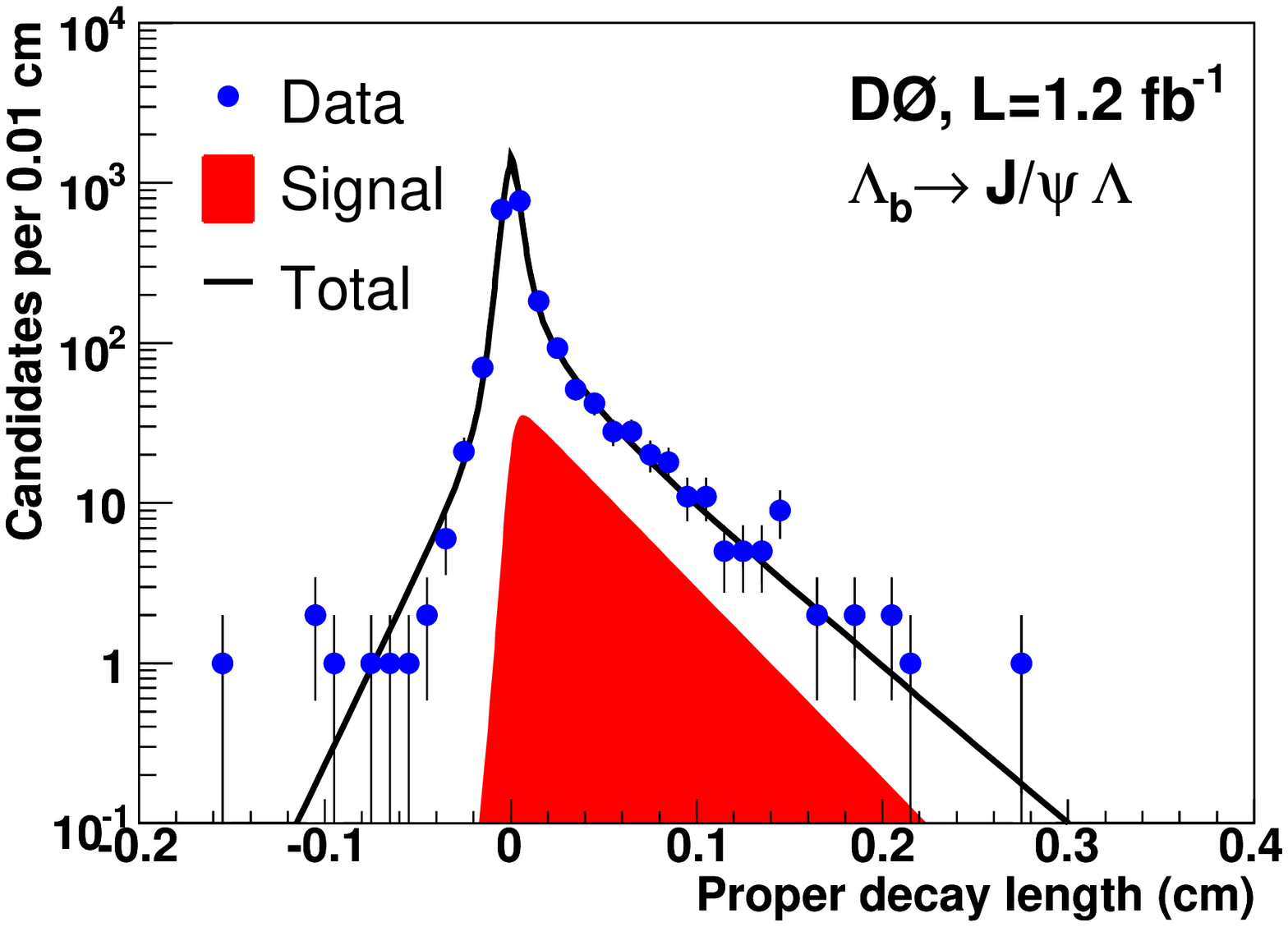}
\caption{Proper decay length distribution of the $\Lambda_b$ candidates from CDF (upper plot) and D\O\ (bottom plot), with the fit result superimposed. 
The shaded regions represent the signal.} 
\label{Lambdab}
\end{figure}

Both CDF and D\O\ have measured the $\Lambda_b$ lifetime in the golden decay mode $\Lambda_b\rightarrow J/\psi\Lambda$. 
Similar analysis procedure have been used by the two experiments, on respectively 1 and 1.2 fb$^{-1}$ of data. . 
The $\Lambda_b$ lifetime was extracted from an unbinned simultaneous likelihood fit to the mass and 
proper decay lenghts distributions. To cross check the validity of the method similar analysis were  
performed on the kinematically similar decay $B^0\rightarrow J/\psi K_s$. Figure~\ref{Lambdab} shows 
the proper decay time distributions of the $J/\psi\Lambda$ pair samples from CDF and D\O. The $\Lambda_b$ lifetime values 
extracted from the maximum likelihood fit to these distributions are~\cite{D0Lb1,CDFLb}: 
$\tau^{\mathrm{CDF}}_{\Lambda_b} = 1.580\pm 0.077(\mathrm{stat})\pm 0.012(\mathrm{sys})$ ps and 
$\tau^{\mathrm{D\O}}_{\Lambda_b}= 1.218^{+0.130}_{-0.115}(\mathrm{stat})\pm0.042(\mathrm{sys})$ ps. \par
The CDF measured value is the single most precise measurement of the $\Lambda_b$ lifetime but is 3.2 
$\sigma$ higher than the current world average~\cite{dmd} ($\tau^{\mathrm{W.A.}}_{\Lambda_b}= 1.230\pm 0.074$ ps). 
The D\O\ result however is consistent with the world average value. The CDF and D\O\ $B^0\rightarrow J/\psi K_s$ measured 
lifetimes are: $\tau^{\mathrm{CDF}}_{B^0} = 1.551\pm 0.019(\mathrm{stat})\pm 0.011(\mathrm{sys})$ ps and 
$\tau^{\mathrm{D\O}}_{B^0} = 1.501^{+0.078}_{-0.074}(\mathrm{stat})\pm0.05(\mathrm{sys})$ ps. Both are compatible 
with the world average value~\cite{HFAG07} ($\tau^{\mathrm{W.A.}}_{B^0}= 1.527\pm 0.008$ ps). One needs more experimental 
input to conclude about the difference between the CDF and the D\O /world average $\Lambda_b$ lifetime values. 
One of the $\Lambda_b$ decay modes that can be exploited is the fully hadronic 
$\Lambda_b\rightarrow \Lambda^+_c\pi^-$, with $\Lambda^+_c\rightarrow pK^-\pi^+$. CDF has in this decay mode 
about 3000 reconstructed events which is 5.6 more than in $\Lambda_b\rightarrow J/\psi\Lambda$. \par 

\begin{figure}[h]
\centering
\includegraphics[width=80mm]{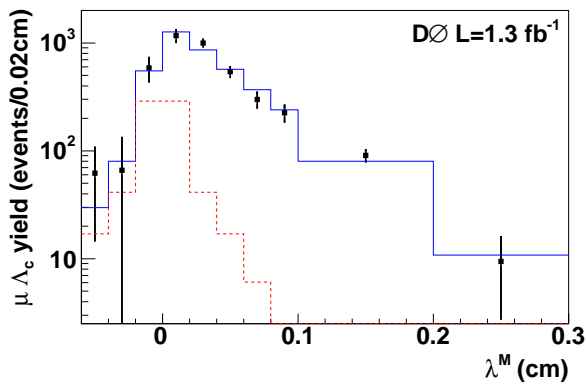}
\caption{Measured yields in the VPDL bins and the result of the lifetime fit. The dashed line shows the $c\bar{c}$ contribution.} 
\label{Lambdacl}
\end{figure}

Recently, the D\O\ experiment has performed a new measurement of the $\Lambda_b$ lieftime in the semileptonic decay 
channel $\Lambda_b\rightarrow \Lambda^+_c\mu^-\bar{\nu}_{\mu}X$, with $\Lambda^+_c\rightarrow K_s p$~\cite{D0Lb2}. 
This measurement is based on 1.2 fb$^{-1}$ of data. As this is a partially reconstructed decay mode the 
proper decay time is corrected by a kinematical factor $K=P_T(\Lambda^+_c\mu^-)/P_T(\Lambda_b)$, estimated from 
Monte Carlo simulation. The $\Lambda_b$ lifetime is not determined from the usually performed unbinned maximum likelihood fit,  
but is extracted from the number of $K_sp\mu^-$ events in bins of their visible proper decay length (VPDL). 
Figure~\ref{Lambdacl} shows the distribution of the number of $\Lambda^+_c\mu^-$ as function of the VPDL with the result of 
the fit superimposed. The fitted $\Lambda_b$ lifetime value is 
$\tau(\Lambda_b)=1.28\pm0.12(\mathrm{stat})\pm0.09(\mathrm{sys})$ ps. This results is compatible with the lifetime value from 
$\Lambda_b\rightarrow J/\psi\Lambda$ and the world average. \par

\begin{acknowledgments}
I would like to thank the local organizer committee for the wonderful and very successful FPCP07 conference.  
\end{acknowledgments}

\bigskip 

\end{document}